\documentclass[preprint,aps,prd]{revtex4}
\usepackage{epsfig}

\RequirePackage{xspace}
\usepackage{epsfig}
\usepackage{graphicx}
\usepackage{color}

\def\mathP{\mathcal{P}}
\def\mathPEW{\mathcal{P}_{EW}}
\def\mathPCEW{\mathcal{P}_{EW}^C}
\def\mathEPCEW{\mathcal{EP}^{C}_{EW}}
\def\mathEP{\mathcal{EP}}
\def\mathT{\mathcal{T}}
\def\mathC{\mathcal{C}}
\def\mathA{\mathcal{A}}

\def\dT{\delta_T}
\def\dC{\delta_C}
\def\dEW{\delta_{EW}}
\def\BR{\mathcal{B}}
\def\ACP{\mathcal{A}_{CP}}

\def\rT{r_T}
\def\rC{r_C}
\def\rEW{r_{EW}}

\def\rCM{r_C^M}
\def\rEWM{r_{EW}^M}
\def\dCM{\dC^M}
\def\dEWM{\dEW^M}

\def\rEWN{r^N}
\def\phiN{\phi^N}
\def\dEWN{\delta^N}

\def\eV{\textrm{eV}}

\def\nn{\nonumber}

\begin{document}

\preprint{}
\title{Analytic Resolution of Puzzle in $B\to K \pi$ decays}
\author{C.~S.~Kim}\email{cskim@yonsei.ac.kr} \affiliation{Department
  of Physics, Yonsei University, Seoul 120-479, Korea}
\author{Sechul~Oh}\email{scoh@phya.yonsei.ac.kr} \affiliation{Department
  of Physics, Yonsei University, Seoul 120-479, Korea}
\author{Yeo Woong Yoon}\email{ywyoon@yonsei.ac.kr}\affiliation{Department
  of Physics, Yonsei University, Seoul 120-479, Korea}

\date{\today}
%
%
\begin{abstract}

\noindent
We present a systematic method to extract each
standard model (SM)-{\it like} hadronic parameter as well as new physics parameters
in analytic way for $B\to K\pi$ decays.
Using the analytic method to the currently available experimental data,
we find two possible solutions analytically equivalent: one showing the
large SM-{\it like} color-suppressed tree contribution and the other showing
the large SM-{\it like} EWP contribution.
The magnitude of the NP amplitude and its weak phase are quite large.
For instance,
we find $|P^{NP}/P| = 0.39\pm0.13$, $\phi^{NP}=91^\circ\pm15^\circ$ and
$\delta^{NP}=8^\circ\pm27^\circ$, which are the ratio of the NP-to-SM
contribution, the weak and the relative
strong phase of the NP amplitude, respectively.
\end{abstract}

\maketitle

\section{Introduction}
\label{sec:1}

Brilliant progress of the $B$ factory experiments sheds light on the
study of rare $B$ decays, which are crucial for testing the standard
model (SM) and detecting any hints beyond the SM.  Especially, $B
\to K \pi$ decays are of great importance not only for investigating
new physics (NP) due to the property of penguin dominance but also
examining one of angles of Cabibbo-Kobayashi-Maskawa (CKM) unitarity
triangle \cite{Nir:1991cu,Fleischer:1997um,Buras:1998rb}.
Many elaborate theoretical calculations based on QCDF~\cite{Beneke:2001ev},
PQCD~\cite{Li:1994iu,Li:2005kt} and SCET~\cite{Bauer:2004tj} have
been done for physical observables within the SM. But, some experimental
data have shown considerable discrepancy from the theoretical estimation,
inspiring searching NP in $B \to K \pi$ decays.

The ratios
\begin{eqnarray}
R_c \equiv 2 \frac{\BR (B^+\to K^+\pi^0)}{\BR (B^+\to K^0\pi^+)} ~~~
{\rm and} ~~~
R_n \equiv \frac{1}{2} \frac{\BR (B^0\to K^+\pi^-)}{\BR (B^0\to K^0\pi^0)}
\end{eqnarray}
are expected to satisfy $R_c \approx R_n$ within the
SM~\cite{Buras:1998rb}. Before ICHEP-2006, those experimental values
had shown a significant discrepancy, but as time passes they were
getting closer to each other~\cite{Fleischer:2007mq}. Current data
updated by March 2007 in HFAG~\cite{Barberio:2006bi} show $R_c =
1.12\pm0.07$ and $R_n = 0.98\pm0.08$, which are consistent with the
SM expectation. On the other hand, the CP asymmetry measurements
still show a disagreement with the SM prediction.  The SM naively
expects $\ACP(B^0\to K^+\pi^-)\approx \ACP(B^+\to K^+\pi^0)$ for the
direct CP asymmetry and $(\sin 2\beta)_{K_S \pi^0}  \approx (\sin
2\beta)_{c \bar c s} =0.68 $ for the mixing-induced CP asymmetry.
But the current experimental data show
\begin{eqnarray}
\label{disc-ACP}
 \ACP(B^+\to K^+\pi^0) &-& \ACP(B^0\to K^+\pi^-) = 0.15 \pm 0.03, \\
\label{disc-S} (\sin 2\beta)_{K_S \pi^0} &-&(\sin 2\beta)_{c \bar c
s} = -0.30 \pm 0.19.
\end{eqnarray}
The recent PQCD result for the difference of the above direct CP asymmetries is
$0.08 \pm 0.09$, which is actually consistent with the data.
However, the PQCD prediction
$\ACP(B^+\to K^+\pi^0)_{PQCD} =-0.01^{+0.03}_{-0.05}$
still has $1.5\sigma$ difference from the current experimental data
$\ACP(B^+\to K^+\pi^0)_{EXP}=0.050\pm0.025$.
Moreover, the difference of the mixing-induced CP asymmetry from the PQCD
prediction is $0.065\pm0.04$, which shows about $2\sigma$ off the data.

Searching for NP via the electroweak penguin (EWP) processes in the
$B \to K \pi$ decays has drawn lots of attention for a long time,
especially based on various specific NP scenarios such as SUSY
models~\cite{Khalil:2005qg}, flavor-changing $Z^{\prime}$
models~\cite{Barger:2004hn}, four generation
models~\cite{Hou:2005hd}, and so on. On the other hand, numerous
model-independent attempts have been also made in search of NP
within the quark diagram approach \cite{Yoshikawa:2003hb,Mishima:2004um,Buras:2003dj,
Buras:2004ub,Baek:2004rp,He:2004ck}.
According to re-parametrization invariance (RI) which was recently
proposed in Ref. \cite{Botella:2005ks,Imbeault:2006nx}, any NP
contribution can be absorbed into the SM amplitudes always in pair:
for example, both the color-suppressed tree and the EWP amplitude.
Thus, we would like to point out that the large enhancement of the
color-suppressed tree amplitude and the EWP amplitude can be
simultaneously understood by the single NP contribution with a
non-zero NP weak phase within the model-independent analysis.

Our main goal in this work is to propose a systematic method for extracting
each hadronic parameters in the presence of the single NP contribution
under the consideration of RI. It will be shown that the parametrization
with the additional NP contribution can be modified into the same form of the
parametrization of the SM. The complete analytic solution for each hadronic
parameters in this SM-{\it like} parametrization will be given in terms
of the experimental data, and also their numerical values.
Therefore, once the experimental data are given, one can pinpoint the hadronic
parameters and will be able to directly compare to the theoretical estimations.
For the extraction of NP parameters, the additional theoretical inputs are
needed. To this end, we adopt two different schemes, one is flavor SU(3) symmetry
and the other is PQCD prediction.
It is discussed that how this NP contribution depends on the weak phase $\gamma$.

\section{Parametrization and Reparametrization Invariance}
\label{sec:2}

In the quark diagram approach~\cite{Gronau:1994bn,Gronau:1995hn},
the decay amplitudes of four $B \to K \pi$ modes are described as
\begin{eqnarray}
\label{pre2-amp0+} A(B^+ \to K^0 \pi^+) &=& \mathP
 + \mathA, \\
\label{pre2-amp+-} A(B^0 \to K^+ \pi^-) &=& -\mathP -\mathPCEW
 - \mathT, \\
\label{pre2-amp+0} \sqrt{2} A(B^+ \to K^+ \pi^0) &=& -\mathP -
\mathPEW - \mathPCEW - \mathT- \mathC -\mathA, \\
\label{pre2-amp00} \sqrt{2} A(B^0 \to K^0 \pi^0) &=& \mathP -
\mathPEW - \mathC\,,
\end{eqnarray}
under the redefinition of
\begin{eqnarray}
\label{def-p-prime} \mathP + \mathEP -
\frac{1}{3} \mathPCEW -\frac{1}{3}\mathEPCEW &\to& \mathP,\\
\label{def-a-prime}\mathA + \mathEPCEW &\to& \mathA \,.
\end{eqnarray}
Each topological parameter represents strong penguin ($\mathP$),
electro-weak penguin ($\mathPEW$), exchange penguin ($\mathEP$), tree
($\mathT$), color-suppressed tree ($\mathC$) and annihilation ($\mathA$)
topologies, respectively.
The superscript $C$ on the penguin parameters denotes a color-suppressed
process.  It is understood that each parameter includes both the weak phase
and the strong phase in it.
Each penguin parameters are involved in three terms associated with the
internal quark exchanges. They can be manipulated by
\begin{eqnarray}
\label{btos-penguin} \mathP
\equiv V^*_{tb}V_{ts}~ \tilde \mathP_{tc} + V^*_{ub}V_{us}~
\tilde \mathP_{uc} \nn
\equiv \mathP_{tc} + \mathP_{uc}~.
\end{eqnarray}
using unitarity of the CKM matrix. Note that the CKM factors relevant to each
parameter are $V^{*}_{tb}V_{ts}$ for the $\mathP_{tc}, \mathP_{EW}, \mathP^{C}_{EW}$
and $V^*_{ub}V_{us}$ for the $\mathT, \mathC, \mathA$, $\mathP_{uc}$.
The relative sizes among these parameters are roughly estimated within the
SM~\cite{Gronau:1995hn} as
\begin{eqnarray}
\label{su3est}
    1&:&|\mathP_{tc}|, \nn \\
\mathcal{O}(\lambda) &:& |\mathT|, |\mathP_{EW}|, \nn \\
\mathcal{O}(\lambda^2) &:& |\mathC|, |\mathP^{C}_{EW}|, \nn \\
\mathcal{O}(\lambda^3) &:& |\mathA|,
\end{eqnarray}
where $\lambda \sim 0.2$ from the Wolfenstein parametrization
\cite{Wolfenstein:1983yz}.
For the relative size of $|\mathP_{uc}|$, one can
roughly estimate that
\begin{equation}
\label{puc-size} \left|\frac{\mathP_{uc}}{\mathP_{tc}}\right| =
\left|\frac{V^*_{ub}V_{us}\tilde \mathP_{uc}}{V^*_{tb}V_{ts}\tilde
\mathP_{tc}}\right| \sim \lambda^2 \left|\frac{\tilde
\mathP_{uc}}{\tilde \mathP_{tc}}\right|.
\end{equation}
Note that $\tilde \mathP_u$ and $\tilde \mathP_c$ are smaller than
$\tilde \mathP_t$~\cite{Baek:2005tj}, and more precisely it can be estimated
that $0.2 < | \tilde \mathP_{uc} / \tilde \mathP_{tc} |< 0.4$ within the
perturbative calculation~\cite{Buras:1994pb}.  Therefore, we assume
$| \mathP_{uc} / \mathP_{tc} | \sim \mathcal{O}(\lambda^3)$ for our analysis.
It has been generally argued that the NP effects, if present, are the size of
the EWP amplitude or smaller in $B \to K \pi$ decays. Thus we neglect all the
minor contributions smaller than $|\mathC|$, such as $\mathA$ and $\mathP_{uc}$,
for simplicity. (We also neglect $\mathP^{ C}_{EW}$, since the
$|\mathP^{C}_{EW}|$ is expected to be smaller than
$|\mathC|$~\cite{Buras:2004ub,Mishima:2004um}.) Therefore, in our analysis the
limit of NP sensitivity  would be the order of $|\mathC| ~(\sim \lambda^2
\mathP_{tc})$ at most.

Explicitly showing the weak phase $\gamma$ and the strong phases $\delta$,
the decay amplitudes can be rewritten as
\begin{eqnarray}
\label{amp0+} A(B^+ \to K^0 \pi^+) &=& -P, \\
\label{amp+-} A(B^0 \to K^+ \pi^-) &=& P(
 1 - \rT e^{i \gamma} e^{i \dT}), \\
\label{amp+0} \sqrt{2} A(B^+ \to K^+ \pi^0) &=& P (1 - \rT e^{i
\gamma}
e^{i \dT} - \rC e^{i \gamma} e^{i \dC} + \rEW e^{i \dEW} ), \\
\label{amp00} \sqrt{2} A(B^0 \to K^0 \pi^0) &=& P (-1 - \rC e^{i
\gamma} e^{i \dC} + \rEW e^{i \dEW}),
\end{eqnarray}
where $P \equiv |\mathP_{tc}|$, $\rT \equiv
|\mathT/\mathP_{tc}|$, $\rC \equiv |\mathC/\mathP_{tc}|$,
$\rEW \equiv |\mathP_{EW}/\mathP_{tc}|$, which are defined to be
positive.  We set the strong phase of the penguin contribution
$P$ to be zero so that all the other strong phases are relative to
it. It is also used that $V^*_{tb} V_{ts} = - |V^*_{tb} V_{ts}|$.
We assume that the weak phase $\gamma$ can be measured from elsewhere.
Then the number of unknown parameters in the above decay amplitudes
is 7 ($P, \rT, \rC, \rEW,$ $\dT, \dC, \dEW$) within the SM. We
again emphasize that this approximated parametrization is the
most efficient way to probe new physics up to the order of
$|\mathC|$.

Now we introduce a single NP contribution coming through the EWP (or the color
suppressed tree) contribution such as
\begin{equation}
\label{EWNPform} P^N e^{i \phiN} e^{i \dEWN},
\end{equation}
where $P^N$ is defined to be positive, and $\phiN$ and $\dEWN$ are weak and
strong phase of the NP term, respectively.
Then the two decay amplitudes in Eqs.~(\ref{amp+0}) and (\ref{amp00}) are modified
by simply adding the NP term in the EWP contribution:
\begin{eqnarray}
\label{amp+0-NP} \sqrt{2} A(B^+ \to K^+ \pi^0) &=& P (1 - \rT
e^{i \gamma} e^{i \dT} - \rC e^{i \gamma} e^{i \dC} + \rEW
e^{i \dEW} +\rEWN e^{i \phiN} e^{i \dEWN}), \\
\label{amp00-NP} \sqrt{2} A(B^0 \to K^0 \pi^0) &=& P (-1 - \rC
e^{i \gamma} e^{i \dC} + \rEW e^{i \dEW}+\rEWN e^{i \phiN} e^{i \dEWN}),
\end{eqnarray}
where $\rEWN \equiv P^N/P$. It has been introduced
that any single decay amplitude can be separated into two decay amplitudes which have
arbitrary weak phases $\theta$ and $\eta$, respectively, unless $\theta$ and $\eta$
are equal or modulo $\pi$~\cite{Botella:2005ks}.
Since any physical results should not be changed, it is called reparametrization
invariance (RI). More explicitly, any phase term $e^{i \phi}$ can be separated as
\begin{equation}
\label{RI}
e^{i \phi} = \frac{\sin(\phi-\eta)}{\sin(\theta-\eta)} e^{i\theta}
 - \frac{\sin(\phi-\theta)}{\sin(\theta-\eta)} e^{i \eta} ~,
\end{equation}
where the phases $\theta$ and $\eta$ are arbitrarily chosen, satisfying
$\theta-\eta \neq 0~ (\textrm{mod}~ \pi)$. This is a simple algebraic identity.
Due to this identity, the NP amplitude can be re-expressed as
\begin{eqnarray}
\label{EW-NP-RI}
\rEWN e^{i \phiN} e^{i \dEWN}
 = \rEWN \frac{\sin \phiN}{\sin\gamma} e^{i \gamma} e^{i \dEWN}
 - \rEWN \frac{\sin (\phiN-\gamma)}{\sin\gamma}e^{i \dEWN}.
\end{eqnarray}
Here, the weak phases $\gamma$ and $0$ are chosen in order to match with the
weak phases of the color-suppressed tree and EWP amplitudes.
Then those two terms can be absorbed into the parameters of
the color-suppressed tree and EWP leading to the following parametrization
\begin{eqnarray}
\label{amp+0-NP2}
\sqrt{2} A(B^+ \to K^+ \pi^0) &=& P (1 - \rT e^{i \gamma} e^{i \dT}
 - r_C^{M} e^{i \gamma} e^{i \dC^{M}} + r_{EW}^{M} e^{i \dEW^{M}}), \\
\label{amp00-NP2}
\sqrt{2} A(B^0 \to K^0 \pi^0) &=& P (-1 - r_C^{M} e^{i \gamma} e^{i \dC^{M}}
 + r_{EW}^{M} e^{i \dEW^{M}}),
\end{eqnarray}
which have the same form of the SM parametrization with the following modified
parameters
\begin{eqnarray}
\label{def-rC-mod}
r_C^{M}e^{i \dC^{M}} &\equiv& \rC e^{i \dC}
 - \rEWN \frac{\sin\phiN}{\sin\gamma}e^{i \dEWN}, \\
\label{def-rEW-mod}
r_{EW}^{M}e^{i \dEW^{M}} &\equiv& \rEW e^{i \dEW}
 - \rEWN \frac{\sin(\phiN-\gamma)} {\sin\gamma}e^{i \dEWN}.
\end{eqnarray}
The NP amplitude is now absorbed into the SM parameters of
the color-suppressed tree and EWP. Therefore,
color-suppressed tree amplitude in these SM-{\it like} parametrization
can be affected by the NP contribution of EWP unless $\phi^N =
0$ as shown in Eq.~(\ref{def-rC-mod}).

\section{Analytic Solutions for the SM-like parameters}
\label{sec:3}
In this section, we present the analytic solutions for the SM-{\it like}
parameters \cite{Kim:2007ee}.
  For the first step, we rewrite Eqs.~(\ref{amp0+}), (\ref{amp+-}),
(\ref{amp+0-NP2}), and (\ref{amp00-NP2}) as
\begin{eqnarray}
\label{Namp0+}
A^{0+} e^{i \alpha^{0+}} &\equiv& A^{0+} e^{i \pi} = -P, \\
\label{Namp+-}
A^{+-} e^{i \alpha^{+-}} &=& P( 1 - \rT e^{i \gamma} e^{i \dT}), \\
\label{Namp+0}
\sqrt{2} A^{+0} e^{i \alpha^{+0}} &=& P (1 - \rT e^{i \gamma} e^{i \dT}
 - r_C^{M} e^{i \gamma} e^{i \dC^{M}} + r_{EW}^{M} e^{i \dEW^{M}}), \\
\label{Namp00}
\sqrt{2} A^{00} e^{i \alpha^{00}} &=& P (-1 - \rCM e^{i \gamma}
 e^{i \dCM} + \rEWM e^{i \dEWM}),
\end{eqnarray}
where $A^{ij}$ denote magnitudes of the decay amplitudes of $B \to K^i \pi^j$ and
$\alpha^{ij}$ represent their complex phases ($ij=\{0+,+-,+0,00\}$).
We put a {\it bar} on top of the amplitude parameters in case of the CP conjugate
modes. It should be noted that these SM-{\it like} parametrization is including
the NP contribution, namely the one coming into EWP sector, via RI.
Table~\ref{table:1} shows current experimental data for the $B \to K
\pi$ decays~\cite{Aubert:2006gm,Bornheim:2003bv}. We use the
notation for the branching ratios and CP asymmetries compatible with
HFAG~\cite{Barberio:2006bi}:
\begin{eqnarray}
\label{BRs}
\BR^{ij} &\propto& \tau_{B^{(+,0)}} \frac{{A^{ij}}^2+\bar{A^{ij}}^2}{2}, \\
\label{ACPs}
\ACP^{ij} &\equiv& -\frac{{A^{ij}}^2-\bar{A^{ij}}^2}{{A^{ij}}^2+\bar{A^{ij}}^2}, \\
\label{S-pipi}
S_f &\equiv& \eta_f \frac{2 \textrm{Im} \lambda_f}{1+|\lambda_f|^2},
\end{eqnarray}
where $\tau_{B^{(+,0)}}$ is the life time of a $B^{(+,0)}$ meson. The $\lambda_f$ is
defined by $\lambda_f=e^{-2 i \beta}\bar{A}/A$
and $\eta_f$ is the CP eigenvalue of the final state $f$.
We also use the following numerical values from PDG~\cite{Yao:2006px}:
\begin{eqnarray}
\sin 2 \beta = 0.687, ~~\gamma=63^\circ,~~\tau_{B^+}/\tau_{B^0}=1.071.
\label{PDGdata}
\end{eqnarray}

\begin{table}[t]
\caption{Current experimental data for $B\to K \pi$. The branching ratios
are in $10^{-6}$.  The average values are given by HFAG, updated by September
2007~\cite{Barberio:2006bi}.}
\smallskip
\begin{tabular}{c|cccc}
\hline
 Measurement  ~~&~~ BABAR  ~~&~~ Belle ~~&~~ CLEO ~~&~~ Average\\
\hline
$\BR(K^0 \pi^+)$ ~&~~$23.9\pm1.1\pm1.0$~&~~$22.8^{+0.8}_{-0.7}\pm1.3$~&~~$18.8^{+3.7+2.1}_{-3.3-1.8}$~&~$23.1\pm1.0$\\
$\BR(K^+ \pi^0)$ ~&~~$13.6\pm0.6\pm0.7$~&~~$12.4\pm0.5\pm0.6$~&~~$12.9^{+2.4+1.2}_{-2.2-1.1}$~&~$12.9\pm0.6$\\
$\BR(K^+ \pi^-)$ ~&~~$19.1\pm0.6\pm0.6$~&~~$19.9\pm0.4\pm0.8$~&~~$18.0^{+2.3+1.2}_{-2.1-0.9}$~&~$19.4\pm0.6$\\
$\BR(K^0 \pi^0)$ ~&~~$10.3\pm0.7\pm0.6$~&~~$9.2\pm+0.7^{+0.6}_{-0.7}$~&~~$12.8^{+4.0+1.7}_{-3.3-1.4}$~~&~$9.9\pm0.6$\\
\hline
$\ACP(K^0 \pi^+)$ ~&~~$-0.029\pm0.039\pm0.010$~&~~$0.03\pm0.03\pm0.01$~&~~$0.18\pm0.24\pm0.02$  ~&~ $0.009\pm0.025$\\
$\ACP(K^+ \pi^0)$ ~&~~$0.030\pm0.039\pm0.010$~&~~$0.07\pm0.03\pm0.01$~&~$-0.29\pm0.23\pm0.02$  ~&~ $0.050\pm0.025$\\
$\ACP(K^+ \pi^-)$ ~&~~$-0.107\pm0.018^{+0.007}_{-0.004}$~&~~$-0.093\pm0.018\pm0.008$~&~~$-0.04\pm0.16\pm0.02$ ~&~ $-0.097\pm0.012$\footnote{This average also includes the CDF result: $-0.086\pm0.023\pm0.009$.}
\\
$\ACP(K^0 \pi^0)$ ~&~~$-0.24\pm0.15\pm0.03$~&~~$-0.05\pm0.14\pm0.05$~&~ ~&~ $-0.14\pm0.11$\\
\hline
$S_{K_S \pi^0}$ ~&~~ $0.40\pm0.23\pm0.03$~&~~$0.33\pm0.35\pm0.08$~&~ ~&~ $0.38\pm0.19$\\
\hline
\end{tabular}
\label{table:1}
\end{table}

The number of parameters is 7~($P, \rT, \rCM, \rEWM, \dT, \dCM, \dEWM$),
while 9 observables are available in $B \to K \pi$ decays.
Since $\ACP^{0+}$ automatically vanishes in our parametrization, we discard the
data.  Setting aside the mixing induced CP asymmetry data
$S_{K_S \pi^0}$, we use the remaining 7 experimental data
in order to determine the 7 parameters.
>From Eq.~(\ref{Namp0+}) we easily get the solution for $P$ in terms of the
observable by taking into account the phase space factor:
\begin{eqnarray}
\label{solP} P = A^{0+} = (49.9\pm1.1) ~\eV.
\end{eqnarray}
Combining Eqs.~(\ref{Namp0+}) and (\ref{Namp+-}), one finds~\cite{Buras:2004ub} that
\begin{eqnarray}
\label{MBR+-} R &=& 1+{\rT}^2 - 2\rT \cos \dT \cos \gamma,\\
\label{MACP+-} -\ACP^{+-}R &=& 2 \rT \sin \dT \sin \gamma,
\end{eqnarray}
where $R$ is given~\cite{Fleischer:1997um} by
\begin{equation}
\label{def-R} R \equiv
\frac{\BR^{+-}}{\BR^{0+}}\frac{\tau_{B^+}}{\tau_{B^0}}=0.90\pm0.05.
\end{equation}
The analytic solutions for $\dT$ and $\rT$ are obtained in terms of the observables
from the above equations are
\begin{eqnarray}
\label{sol-cotdT} \cot\dT &=& \frac{\sin
2\gamma}{(-\ACP^{+-})R}\left[1 \pm
\sqrt{1+\frac{1}{\cos^2\gamma}\left(R-1-\left(\frac{-\ACP^{+-}R}{2
\sin \gamma}\right)^2 \right)}~\right], \\
\label{sol-rT} \rT &=& \sqrt{R\left(1-\ACP^{+-} \cot \gamma \cot
\dT\right)-1}~.
\end{eqnarray}

\begin{figure}[t]
\centerline{\epsfig{figure=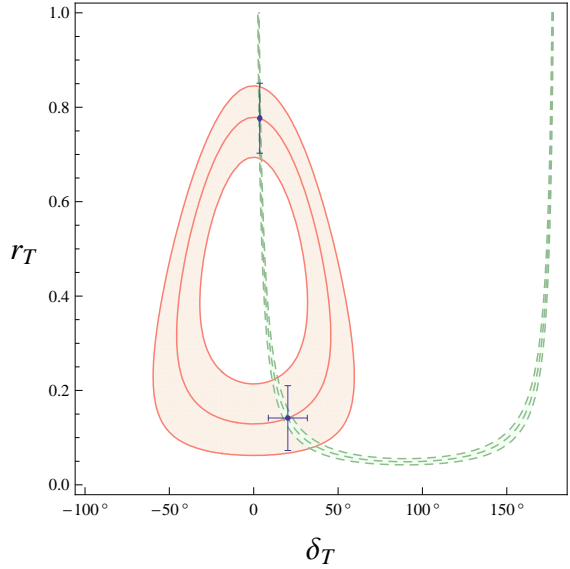, scale=0.7}} \caption{Contour
plot corresponding to the $1 \sigma$ range of $R$ and $\ACP^{+-}$ in
Eqs.~(\ref{MBR+-}) and (\ref{MACP+-}) in the $\rT$-$\dT$ plane.
The solid lines are from Eq.~(\ref{MBR+-}) and the dashed lines are
from Eq.~(\ref{MACP+-}). The two intersection regions show the two
different solutions for $\rT$ and $\dT$.  The two solutions are
marked with error bars. } \label{fig1}
\end{figure}

Using the experimental data given in Table~\ref{table:1}, we obtain numerical values
of $\rT$ and $\dT$.  As shown in Fig.~\ref{fig1}, the following two solutions are
found:
\begin{eqnarray}
\label{sol1-rT-delT} \rT &=& 0.14 \pm 0.07, ~~~~ \dT = 20^\circ
\pm 12^\circ ~, \\
\label{sol2-rT-delT} \textrm{or}~~ \rT &=& 0.78 \pm 0.07,~~~~ \dT
= 3.6^\circ \pm0.5^\circ ~.
\end{eqnarray}
Since the second value of $\rT$ is unreasonably larger than the SM
expectation, which is around 0.15, we safely choose the first one as
our solution.

 The next step is to determine $\alpha^{00}$ and $\bar \alpha^{00}$ in
terms of the experimental data.
After subtracting Eq.~(\ref{Namp00}) from Eq.~(\ref{Namp+0})
and also considering their CP conjugate modes, we get the following equations:
\begin{eqnarray}
\label{Amp+--Amp00}
\sqrt{2} \left(A^{+0} e^{i \alpha^{+0}} - A^{00} e^{i \alpha^{00}}\right)
 = P x e^{i \zeta}~, \\
\label{Amp+--Amp00bar}
\sqrt{2} \left(\bar{A}^{+0} e^{i \bar{\alpha}^{+0}}
 - \bar{A}^{00} e^{i \bar{\alpha}^{00}}\right) = P \bar{x} e^{i \bar \zeta}~,
\end{eqnarray}
where
\begin{eqnarray}
\label{def-x-zeta}
x e^{i \zeta} &\equiv& 2 - r_T e^{i \gamma} e^{i \dT} ~,\\
\label{def-x-zeta2}
\bar x e^{i \bar \zeta} &\equiv& 2 - r_T e^{-i \gamma} e^{i \dT} ~.
\end{eqnarray}
It is easy to find $\alpha^{00}$ and $\bar \alpha^{00}$ from these equations:
\begin{eqnarray}
\label{sol-alpha00} \alpha^{00}&=& \zeta \pm \textrm{ArcCos} \left(
\frac{2 {A^{+0}}^2-2{A^{00}}^2-P^{2}
x^2}{2\sqrt{2}A^{00} P x}\right), \\
\label{sol-alphabar00} \bar \alpha^{00}&=& \bar \zeta \pm
\textrm{ArcCos} \left( \frac{2 \bar {A^{+0}}^2 - 2 \bar
{A^{00}}^2-P^{2} \bar x^2}{2\sqrt{2}\bar A^{00} P \bar
x}\right).
\end{eqnarray}
There occurs a two-fold ambiguity for $\alpha^{00}$ and also for
$\bar \alpha^{00}$. We call them $[\alpha^{00}_{(1)}$,
$\alpha^{00}_{(2)}]$ and $[\bar \alpha^{00}_{(1)}$, $\bar
\alpha^{00}_{(2)}]$, respectively. Consequently, the solution for
$\alpha^{00}$ and $\bar \alpha^{00}$ has a four-fold ambiguity in
total due to their combinations. For convenience, we represent
each case as {\it Case 1, 2, 3}, and {\it 4}, respectively,
corresponding to the combinations of $\left( \alpha^{00}_{(1)},~
\bar\alpha^{00}_{(1)} \right)$, $\left( \alpha^{00}_{(1)},~
\bar\alpha^{00}_{(2)} \right)$, $\left( \alpha^{00}_{(2)},~
\bar\alpha^{00}_{(1)} \right)$, and $\left( \alpha^{00}_{(2)},~
\bar\alpha^{00}_{(2)} \right)$. However, in reality for any given
$\alpha^{00}_{(1,~2)}$ (or $\bar \alpha^{00}_{(1,~2)}$), there
exist only two possible cases: for instance, for given
$\alpha^{00}_{(1)}$, only {\it Case 1} and {\it Case 2} are
possible solutions which indicates a two-fold ambiguity.

\begin{figure}[t]
\centerline{\epsfig{figure=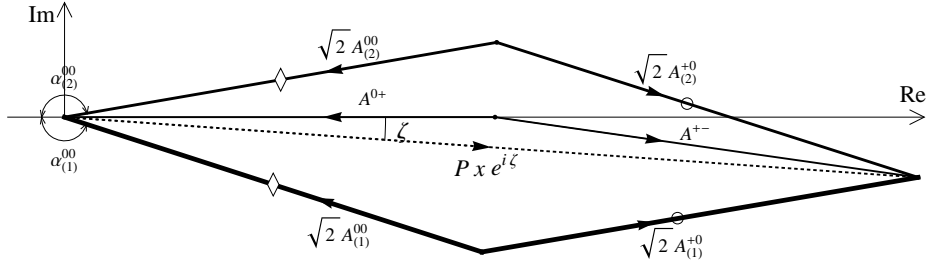, scale=0.6}}
\caption{The isospin quadrangles in a complex plane displaying the isospin
relation among the decay amplitudes for $B \to K\pi$.
$A^{ij}_{(1,~2)}$ corresponds to the case of $\alpha^{00}_{(1,~2)}$. }
\label{fig2}
\end{figure}

It is instructive to represent the phases $\alpha^{00}$ and $\bar \alpha^{00}$
geometrically as in Fig.~\ref{fig2}.  The figure shows the famous isospin quadrangle
in a complex plane depicting the isospin relation among the decay amplitudes for
$B \to K\pi$:
\begin{eqnarray}
A(B^+ \to K^0 \pi^+) + \sqrt{2} A(B^+ \to K^+ \pi^0)
 = A(B^0 \to K^+ \pi^-) + \sqrt{2} A(B^0 \to K^0 \pi^0) ~.
\label{isospin_rel}
\end{eqnarray}
The notation $A(B \to K^i \pi^j) \equiv A^{ij} e^{i\alpha^{ij}}$ is used in the figure
and $A^{ij}_{(1,~2)}$ corresponds to the case of $\alpha^{00}_{(1,~2)}$.
The isospin quadrangle can be geometrically constructed as follows.
The two complex values of  $A(B^+ \to K^0 \pi^+)$ and $A(B^0 \to K^+ \pi^-)$
in the complex plane are fixed from the solutions shown above.
Subsequently, the
value $x e^{i\zeta}$ is determined, where $x e^{i\zeta} \equiv A(B^0 \to K^+ \pi^-)
- A(B^+ \to K^0 \pi^+) = \sqrt{2} A(B^+ \to K^+ \pi^0) -\sqrt{2} A(B^0 \to K^0 \pi^0)$
as defined in Eq.~(\ref{def-x-zeta}). Since the magnitudes $A^{+0}$ and $A^{00}$ are
directly determined from the measurements, we find two distinct solutions for
$A(B^+ \to K^+ \pi^0)$ and $A(B^0 \to K^0 \pi^0)$ which are expressed as
$A^{+0}_{(1,~2)}$ and $A^{00}_{(1,~2)}$ in Fig.~\ref{fig2}.
Two sides of the quadrangles denoted by the diamond marks (and the circle marks) are
equal in length to each other.
The quadrangles in the figure have been constructed by using the present experimental
data. We recall that the weak phase $\gamma$ has been used as an input in
Eq.~(\ref{PDGdata}).

 Next, we find analytic solutions for $\rCM, \rEWM, \dCM, \dEWM$ in terms
of the observables, using $\alpha^{00}$ and $\bar \alpha^{00}$ determined in
Eqs.~(\ref{sol-alpha00}) and (\ref{sol-alphabar00}).
To this end, we use Eq.~(\ref{Namp00}) and its CP conjugate version.  They can be
rewritten as
\begin{eqnarray}
\label{Eq-rC-rEW}
- \rCM e^{i \gamma} e^{i \dCM} + \rEWM e^{i \dEWM} = y e^{i \eta}, \\
\label{Eq-rC-rEW-2}
- \rCM e^{-i \gamma} e^{i \dCM} + \rEWM e^{i \dEWM} = \bar y e^{i \bar \eta},
\end{eqnarray}
where
\begin{eqnarray}
\label{def-y}
y e^{i \eta} &\equiv& \sqrt{2} \frac {A^{00}}{P}e^{i \alpha^{00}} +1 ~, \\
\label{def-ybar}
\bar y e^{i \bar \eta} &\equiv& \sqrt{2} \frac
 {\bar A^{00}}{P}e^{i \bar \alpha^{00}} +1 ~.
\end{eqnarray}
\begin{table}[b]
\caption{Four possible solutions for $\rCM,~ \rEWM,~ \dCM,~ \dEWM$
 and the prediction for $S_{K_S \pi^0}$ in each case. Current experimental value
 for  $S_{K_S \pi^0}$ is $0.38\pm0.19$.
 }
\smallskip
\begin{tabular}{cccccc}
\hline
  &  $\rCM$ & $\rEWM$ & $\dCM$ & $\dEWM$ & $S_{K_S \pi^0}$\\
\hline
{\it Case 1}~~&~~$0.085 \pm 0.080$ ~~&~~ $0.25 \pm 0.11$ ~~&~~
 $226^\circ \pm 81^\circ$ ~~&~~ $77^\circ \pm 16^\circ$ ~~&~~$0.69 \pm 0.14$\\
{\it Case 2}~~&~~$0.36 \pm 0.13$ ~~&~~ $0.068 \pm 0.064$ ~~&~~
 $192^\circ \pm 11^\circ$ ~~&~~ $202^\circ \pm 78^\circ$ ~~&~~$0.08 \pm 0.26$\\
{\it Case 3}~~&~~$0.24\pm 0.13$ ~~&~~ $0.17 \pm 0.10$ ~~&~~
 $-20^\circ \pm 15^\circ$ ~~&~~ $-6.4^\circ \pm 26^\circ$ ~~&~~$0.92 \pm 0.07$\\
{\it Case 4}~~&~~$0.12\pm 0.11$ ~~&~~ $0.29\pm0.13$ ~~&~~
 $235^\circ \pm 45^\circ$ ~~&~~ $-80^\circ \pm 15^\circ$ ~~&~~$0.55 \pm 0.16$\\
\hline
\end{tabular}
\label{table:2}
\end{table}
It is straightforward  to find the
solutions for $r_C^{M},~ r_{EW}^{M},~ \dC^{M},~ \dEW^{M}$ as a function of
$y,~ \bar y$ and $\eta,~ \bar \eta$ from Eqs.~(\ref{Eq-rC-rEW}) and (\ref{Eq-rC-rEW-2}) :
\begin{eqnarray}
\label{sol-rC} \rCM &=& \frac{1}{2 \sin{\gamma}} \sqrt{
 |y|^2 + |\bar y|^2 -2 y \bar y \cos(\bar \eta -\eta)},\\
\label{sol-rEW} \rEWM &=& \frac{1}{2 \sin{\gamma}} \sqrt{ |y|^2 +
|\bar y|^2 -2 y \bar y \cos(2 \gamma +\bar \eta
-\eta)}, \\
\label{sol-delC} \dCM &=& \textrm{ArcTan} \left( -\frac{y
\cos\eta-\bar y \cos \bar \eta}{y \sin \eta -\bar y \sin \bar
\eta } \right), \\
\label{sol-delEW} \dEWM &=& \textrm{ArcTan} \left( -\frac{y
\cos(\eta-\gamma)-\bar y \cos(\bar \eta +\gamma)}{y
\sin(\eta-\gamma)-\bar y \sin(\bar \eta +\gamma)} \right).
\end{eqnarray}
We note that there occurs no ambiguity in the above equations.
Therefore, we have
found the analytic solutions for the 7 parameters: $(P,~ \rT,~
\dT)$ without ambiguity, and $(\rCM,~ \rEWM,~ \dCM,~ \dEWM)$ with a
four-fold discrete ambiguity which stems from $\alpha^{00}$ and $\bar \alpha^{00}$ given in
(\ref{sol-alpha00}) and (\ref{sol-alphabar00}).

Now we substitute the values of experimental data into our analytic
solutions in order to get the numerical values of $\rCM,~ \rEWM,~
\dCM$, and $\dEWM$. Table~\ref{table:2} shows the result for each
case.  The prediction for $S_{K_S\pi^0}$ is also given for each case.
Provided precise measurement of $S_{K_S\pi^0}$,
one can choose consistent solutions with the data
of $S_{K_S\pi^0}$ among these 4 cases. Then, as mentioned before,
one can analyze each hadronic parameters of the solutions,
comparing to given theoretical estimation such as PQCD and QCDF.
The {\it Case 3} solution is discarded because its prediction for
$S_{K_S\pi^0}$ is quite different from the current data.
The solutions for {\it Cases 2} and {\it 4} are our favorites because their
predictions for $S_{K_S\pi^0}$ are consistent with the data within
$1\sigma$ error. The {\it Case 2} solution shows large color-suppressed tree
than the typical SM estimation, while the {\it Case 4} solution presents large EWP,
where both cases suggest considerable NP contribution.

Please note that many authors uncovered that the
anomalous behaviors of the experimental data could be accommodated
with the enhancement of the EWP
amplitude~\cite{Yoshikawa:2003hb,Mishima:2004um} as well as an
additional weak phase in the electroweak
sector~\cite{Buras:2003dj,Buras:2004ub,Baek:2004rp}, and a few authors
have also found that the color-suppressed tree amplitude would be
the main source of NP in the $B \to K \pi$ modes~\cite{Baek:2004rp,
He:2004ck}.
Due to our analytic approach, we can find
two solutions analytically equivalent: one showing the
large SM-{\it like} color-suppressed tree contribution and the other showing
the large SM-{\it like} EWP contribution.

\section{Extracting New Physics Parameters and Discussion }
\label{sec:4}

 Finally, we would like to solve Eqs.~(\ref{def-rC-mod}) and (\ref{def-rEW-mod})
for the NP parameters $\rEWN,~ \dEWN$ and $\phiN$.
The left-hand side of Eqs.~(\ref{def-rC-mod}) and (\ref{def-rEW-mod})
has 4 parameters which can be obtained from the analytic solution shown above.
But the number of unknown parameters on the right-hand side
is 7 ($r_C,~ \dC,~ r_{EW},~ \dEW,~ \rEWN,~ \dEWN,~ \phiN$)
Thus there is no model independent way to extract NP parameters without additional
theoretical inputs. We need at least 3 additional inputs in the color-suppressed
tree and the EWP sector in order to determine NP parameters.
Here, we adopt two different schemes for the additional theoretical inputs:
one is flavor SU(3) symmetry, and the other is recent PQCD calculation.

 Using the flavor SU(3) symmetry, we estimate the color-suppressed tree
amplitude from the $B\to\pi\pi$ decays amplitudes,
following the Ref. \cite{Buras:2004ub},
\begin{eqnarray}
\label{pC-SU3} C = \frac{\lambda}{1-\lambda^2/2} ~C_{\pi\pi} ~~ &=& (3.8\pm0.4) ~\eV, \\
\label{dC-SU3} \dC &=& -12^\circ\pm15^\circ\,,
\end{eqnarray}
where the $C_{\pi\pi}$ is color-suppressed tree amplitude of $B\to\pi\pi$ decays.
The EWP amplitude is also associated with
the tree and color-suppressed amplitudes under flavor SU(3)
symmetry \cite{Neubert:1998pt} as
\begin{equation}
\label{pEWdEW-SU3} \rEW e^{i \dEW} =
-\frac{3}{2}\frac{c_9+c_{10}}{c_1+c_2}\frac{1}{\lambda^2 R_b}(\rT
e^{i\dT}+\rC e^{i\dC})~.
\end{equation}
The parameters $\rT$ and $\dT$ can be obtained within the $B\to K\pi$ modes as in
Eq.~(\ref{sol1-rT-delT}).  And the parameters $\rC$ and $\dC$ are given by
Eqs.~(\ref{pC-SU3}) and (\ref{dC-SU3}) combined with Eq.~(\ref{solP}).  Subsequently
$\rEW$ and $\dEW$ can be obtained from the above equation. We summarize the result:
\begin{eqnarray}
\label{rCrEW-SU3}
\begin{array}{l}
\rC e^{i\dC} = (0.076\pm0.008)~e^{i(-12\pm15)^\circ}, \\
\rEW e^{i\dEW} = (0.14\pm0.04)~e^{i(9\pm10)^\circ}.
\end{array}
~ ~ ~ ~ ~ ~ \textrm{SU(3)}
\end{eqnarray}
On the other hand,
The Recent PQCD calculation for the $B \to K\pi$ decays
gives \cite{Li:2005kt}
\begin{eqnarray}
\label{rCrEW-PQCD}
\begin{array}{l}
\rC e^{i\dC} = (0.039)~e^{-i 61^\circ}, \\
\label{pEWdEW-PQCD} \rEW e^{i\dEW} = (0.12)~e^{i 22^\circ}.
\end{array}
~ ~ ~ ~ ~ ~ \textrm{PQCD}
\end{eqnarray}
We use these two different schemes for the values of SM parameters in order to
extract NP parameters. Actually, only 3 additional inputs are enough to extract
the NP parameters. Nevertheless, we adopt above 4 additional inputs in order to
get rid of discrete ambiguity.
\begin{table}[t]
\caption{Numerical values of the new physics parameters after using the additional
inputs of the SM parameters from the flavor SU(3) symmetry and PQCD result,
respectively. The result is shown for the {\it Case 2} and {\it Case 4}.}
{\begin{tabular}{@{}ccccc@{}}
\hline
  &   & $r^N$ & $\phi^N$ & $\delta^N$  \\
\hline
SU(3) symmetry
& $\begin{array}{c} \textrm{{\it Case 2}} \\ \textrm{{\it Case 4}} \end{array}$
& $\begin{array}{c} 0.39\pm0.13 \\ 0.29 \pm 0.19 \end{array}$
& $\begin{array}{c} 91^\circ\pm15^\circ \\ 150^\circ\pm24^\circ  \end{array}$
& $\begin{array}{c} 8^\circ\pm27^\circ \\ 29^\circ\pm17^\circ  \end{array}$
\\
\hline
PQCD
& $\begin{array}{c} \textrm{{\it Case 2}} \\ \textrm{{\it Case 4}} \end{array}$
& $\begin{array}{c} 0.34\pm0.13 \\ 0.31 \pm 0.30 \end{array}$
& $\begin{array}{c} 93^\circ\pm15^\circ \\ 162^\circ\pm21^\circ  \end{array}$
& $\begin{array}{c} 7^\circ\pm28^\circ \\ 36^\circ\pm14^\circ  \end{array}$
\\
\hline
\end{tabular}}
\end{table}

  We define the following quantities:
\begin{eqnarray}
\label{delta-rC}
\Delta r_C e^{i \Delta \dC} &\equiv& \rCM e^{i\dCM}-\rC e^{i\dC},\\
\label{delta-rEW}
\Delta r_{EW} e^{i \Delta \dEW} &\equiv& \rEWM e^{i\dEWM}-\rEW e^{i\dEW}.
\end{eqnarray}
The parameters of $\Delta r_C$, $\Delta \dC$, $\Delta r_{EW}$, and $\Delta \dEW$
can be extracted using above additional theoretical inputs. Then,
we can easily see from Eqs.~(\ref{def-rC-mod}) and (\ref{def-rEW-mod}) that the
following relation should be satisfied:
\begin{eqnarray}
\label{sol-test}
\Delta \dC = \Delta \dEW ~(\textrm{mod}~ \pi)=\dEWN ~(\textrm{mod}~ \pi).
\end{eqnarray}
And, we find the solutions of NP parameters as
\begin{eqnarray}
\label{eq-dEWN}
\dEWN &=& \Delta \dEW ~~\textrm{or}~~\Delta \dEW - \pi, \\
\label{eq-phiN}
\frac{\sin \phi^N}{\sin (\phi^N-\gamma)} &=&  \frac{\Delta r_C}{\Delta r_{EW}}, \\
\label{eq-rEWN}
\rEWN &=& \frac{\sin\gamma}{\sin\phi^N} \Delta r_C.
\end{eqnarray}
For the $\dEWN$, two different solutions are
possible as shown in Eq.~(\ref{eq-dEWN}). Since the strong phase of
 NP contribution is expected to be small, we choose the one with
close to the $\dEW$.
The numerical values for the solution with the two different schemes of theoretical
inputs are shown in Table III. We can see that the result is consistent
each other for both schemes of theoretical input.
Note that in both cases, for both schemes of theoretical input,
the magnitude of the NP amplitude is quite large and its
weak phase is also sizable.

\begin{figure}[t]
\centerline{\epsfig{figure=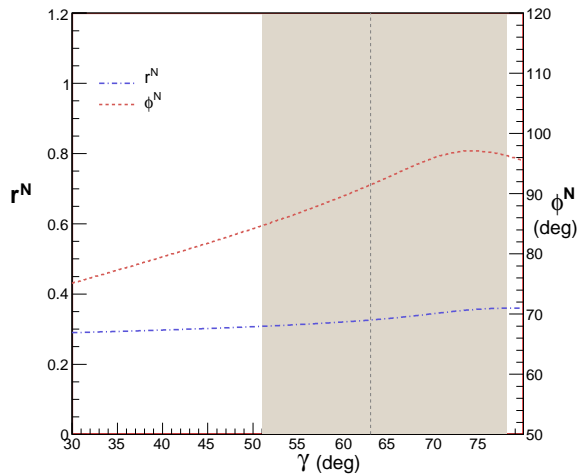, scale=0.4}}
\caption{The $\chi^2$ fitting result for the new physics parameters $r^N$
and $\phi^N$ as a function of $\gamma$, using the SU(3) symmetry input.
The shaded area is the experimentally allowed region of $\gamma$ given
in PDG 2006.}
\label{fig3}
\end{figure}

 Since the experimental value of $\gamma$ still has large
uncertainties, we investigate how our NP solutions depend on these experimental
results. We perform a minimum $\chi^2$ analysis to get the NP solution in order to
simply see the dependence. After employing four additional inputs of $\rC, ~\dC,
~\rEW, ~\dEW$ from flavor SU(3) symmetry, the number of unknown parameters is
6 ($P, ~\rT, ~\dT, ~\rEWN, ~\phiN, ~\dEWN$) while we can use 8 available
experimental data excluding $\ACP^{0+}$. The fitting result as a function of
$\gamma$ is shown in Fig.~\ref{fig3}. As we can see, the NP contribution is not much
sensitive to $\gamma$.


\section{Conclusions}
\label{sec:6}
 In this work, we present complete analytic method for analyzing the hadronic
parameters with the single NP contribution under consideration of
reparametrization invariance.
It is shown that any single NP contribution
in the color-suppressed tree sector or EWP sector can affect both the SM
parameters of color-suppressed tree and EWP.
We show the analytic solution for every
parameters of SM-{\it like} parametrization, and also for the NP
parameters. Therefore one can pinpoint each hadronic parameters and compare
them to the theoretical estimations once the precise experimental data are given.
There were 4 possible solutions for the SM-{\it like} parameters
which can be chosen rightfully by considering mixing induced CP asymmetry data.
Consequently, it could be understood simultaneously that the two different
intriguing solutions occur: one is large color-suppressed tree and the other
is large EWP.
We obtain the solution for the NP parameters after adopting additional theoretical
input. The solution shows quite large NP contribution and sizable weak phase of it.
\\

\vspace{1cm}
\centerline{\bf ACKNOWLEDGMENTS}
\noindent
The work of C.S.K. was supported in part by CHEP-SRC and in part by the KRF
Grant funded by the Korean Government (MOEHRD) No. KRF-2005-070-C00030.
The work of S.O. was supported by the Second Stage of Brain Korea 21 Project.
The work of Y.W.Y. was supported by the KRF Grant funded by the Korean Government
(MOEHRD) No. KRF-2005-070-C00030.
\\


\appendix

\end{document}